\documentclass[a4paper,10pt,twoside]{cpc-hepnp}
\usepackage{CJK,upgreek,fancyhdr}
\usepackage{multicol}
\usepackage{graphicx}
\usepackage{booktabs}
\usepackage{amssymb,bm,mathrsfs,bbm,amscd}
\usepackage[tbtags]{amsmath}
\usepackage{lastpage}

\begin{document}
\begin{CJK*}{GBK}{song}

\fancyhead[c]{\small Chinese Physics C~~~Vol. 41, No. 4 (2017) 043105}
\fancyfoot[C]{\small 043105-\thepage}

\footnotetext[0]{Received xx November 2016}

\title{Analytic solutions in the acoustic black hole analogue of the conical Kerr metric\thanks{Supported by Conselho Nacional de Desenvolvimento Cient\'{i}fico e Tecnol\'{o}gico (140612/2014-9) }}

\author{%
      H. S. Vieira$^{1,2;1)}$\email{horacio.santana.vieira@hotmail.com}%
}
\maketitle

\address{%
$^1$ Departamento de F\'{i}sica, Universidade Federal da Para\'{i}ba, Caixa Postal 5008, CEP 58051-970, Jo\~{a}o Pessoa, PB, Brazil\\
$^2$ Centro de Ci\^{e}ncias, Tecnologia e Sa\'{u}de, Universidade Estadual da Para\'{i}ba, CEP 58233-000, Araruna, PB, Brazil\\
}

\begin{abstract}
We study the sound perturbation of a rotating acoustic black hole in the presence of a disclination. The radial part of the massless Klein-Gordon equation is written into a Heun form, and its analytical solution is obtained. These solutions have an explicit dependence on the parameter of the disclination. We obtain the exact Hawking-Unruh radiation spectrum.
\end{abstract}



\footnotetext[0]{\hspace*{-3mm}\raisebox{0.3ex}{$\scriptstyle\copyright$}2013
Chinese Physical Society and the Institute of High Energy Physics
of the Chinese Academy of Sciences and the Institute
of Modern Physics of the Chinese Academy of Sciences and IOP Publishing Ltd}%

\begin{multicols}{2}

%
%
\section{Introduction}
Investigations concerning the behavior of scalar fields around black holes may reveal some aspects of their internal structure as well as give us some relevant informations about the physics of these objects \cite{PhysLettB.686.279,PhysRevD.84.127502,PhysRevD.87.084047,ClassQuantumGrav.31.045003}, among which we can highlight the spontaneous emission of black body radiation by black holes: the Hawking radiation \cite{CommunMathPhys.43.199}. The existence of an event horizon in the hydrodynamic analogue suggests that this interesting phenomenon can be produced, in principle, to experimentally verify the analogue Hawking radiation emitted by acoustic black holes \cite{PhysRevD.82.044013,PhysLettB.692.61,PhysLettB.697.398,EurophysLett.109.60006}.

There are many features in common between astrophysical black holes and the corresponding analogue gravity models, although they have different dynamics. For example, the dynamics of astrophysical black holes are governed by Einstein's equations, while the dynamics of acoustic black holes are given by the equations of fluid mechanics. In fact, these different types of black holes possess some fundamental properties in common, which has motivated a lot of investigation into the physics of acoustic black holes \cite{PhysRevD.83.124016,PhysRevD.83.124047,NewJPhys.10.103001,PhysRevD.92.024043,ClassQuantumGrav.18.1137,GenRelGrav.34.1719,JHighEnergyPhys.04.030,JPhysBAtMolOptPhys.45.163001,PhysRevA.78.021601,PhysRevD.90.104015,PhysRevD.81.124013,PhysRevD.79.064014,PhysRevD.70.124006,PhysRevD.44.1731,PhysRevD.48.728}.

A cosmic string can appear as part of a larger gravitational system, for instance, passing through a black hole. In general, solutions of Einstein's equations can be constructed by including a cosmic string \cite{Vilenkin:1994}. According to Gal'tsov and Mas\'{a}r \cite{ClassQuantumGrav.6.1313}, the conical structure of spacetime gives rise to global effects which can be used, in principle, to detect cosmic strings. In this way, some of these effects can be revealed within the background of the Kerr metric with the additional contribution of a string lying along the polar axis.

The interest in the study of these structures, and more specifically the role played by their topological properties in quantum systems, is justified by the richness of the new ideas that they bring to the general relativity. Futhermore, there exist some interesting gravitational effects associated with the non-trivial topology of the spacelike section arround the cosmic string. For example, a cosmic string can act as a gravitational lens, it can induce a finite electrostatic self-force on an electrically charged particle and an emission of radiation by a freely moving particle. In this way, we can construct an analogue model for the Kerr spacetime with a cosmic string in which the disclination \cite{Marques:2010} mimics the cosmic string. The disclination can amplify or reduce the Hawking-Unruh radiation emitted in agreement with the value of the deficit angle.

The organization of this paper is as follows. In Section 2 we introduce the metric that corresponds to the Kerr black hole with a cosmic string and we present the solutions of the Klein-Gordon equation for a uncharged massive scalar field. In Section 3 we introduce the metric that corresponds to the draining bathtub fluid flow with a disclination, using a geometric procedure to introduce this topological defect. In Section 4 we write down the covariant Klein-Gordon equation for a massless scalar field in the background under consideration, and in Section 5 we determine the Hawking radiation spectrum. Finally, we conclude in Section 6.
%
%
\section{Kerr spacetime with a cosmic string}
In a recent paper \cite{AnnPhys.362.576}, we have analyzed the interaction between scalar fields and the gravitational and electromagnetic field produced by a Kerr-Newman-Kasuya black hole with a cosmic string along its axis of symmetry, whose spacetime is described by the following metric
\begin{eqnarray}
ds^{2} & = & -\frac{\Delta-a^{2}\sin^{2}\theta}{\rho^{2}}\ dt^{2}+\frac{\rho^{2}}{\Delta}\ dr^{2}+\rho^{2}\ d\theta^{2}\nonumber\\
& + & \frac{(r^{2}+a^{2})^{2}-\Delta a^{2}\sin^{2}\theta}{\rho^{2}}b^{2}\sin^{2}\theta\ d\phi^{2}\nonumber\\
& - & 2\frac{(r^{2}+a^{2})-\Delta}{\rho^{2}}a b\sin^{2}\theta\ dt\ d\phi\ ,
\label{eq:metrica_Kerr-Newman-Kasuya_string}
\end{eqnarray}
with
\begin{equation}
\Delta=r^{2}-2Mr+a^{2}+Q_{e}^{2}+Q_{m}^{2}=(r-r_{+})(r-r_{-})\ ,
\label{eq:Delta_metrica_Kerr-Newman-Kasuya}
\end{equation}
\begin{equation}
r_{\pm}=M \pm [M^{2}-(a^{2}+Q_{e}^{2}+Q_{m}^{2})]^{\frac{1}{2}}\ ,
\label{eq:sol_padrao_Kerr-Newman-Kasuya}
\end{equation}
\begin{equation}
\rho^{2}=r^{2}+a^{2}\cos^{2}\theta\ ,
\label{eq:rho_metrica_Kerr-Newman-Kasuya}
\end{equation}
where $a=J/M$, $Q_{e}$, $Q_{m}$, and $M$ are the angular momentum per mass, electric charge, magnetic charge, and mass (energy), respectively. The parameter $b$ codifies the presence of the cosmic string, assuming values in the interval $0 < b < 1$. Indeed, for the special case when $Q_{e}=0=Q_{m}$, Eq.~(\ref{eq:metrica_Kerr-Newman-Kasuya_string}) becomes the stationary spacetime containing a cosmic string which lies along the symmetry axis, i.e., the conical Kerr metric in Boyer-Lindquist coordinates obtained by Gal'tsov and Mas\'{a}r.

Therefore, in this case, the general exact solutions for both angular and radial parts of the Klein-Gordon equation for an uncharged massless scalar field in Kerr spacetime with a cosmic string is obtained from Ref.~\cite{AnnPhys.362.576}, putting $Q_{e}=Q_{m}=0$ and $e=0$ in their Eqs.~(68)-(73) and Eqs.~(82)-(87), respectively. They can be written as
\begin{eqnarray}
W(z) & = & \mbox{e}^{\frac{1}{2}\alpha z}z^{\frac{1}{2}\beta}(z-1)^{\frac{1}{2}\gamma}\nonumber\\
& \times & \{C_{1}\ \mbox{HeunC}(\alpha,\beta,\gamma,\delta,\eta;z)\nonumber\\
& + & C_{2}\ z^{-\beta}\ \mbox{HeunC}(\alpha,-\beta,\gamma,\delta,\eta;z)\}\ ,
\label{eq:solucao_geral_angular_Kerr_z}
\end{eqnarray}
where $C_{1}$ and $C_{2}$ are constants, $\mbox{HeunC}(\alpha,\beta,\gamma,\delta,\eta;z)$ are the confluent Heun functions \cite{Ronveaux:1995,Slavyanov:2000}, and the parameters $\alpha$, $\beta$, $\gamma$, $\delta$, and $\eta$ are shown in Table \ref{tab:Parameters_conical_Kerr}.

\end{multicols}

\begin{center}
\tabcaption{ \label{tab:Parameters_conical_Kerr}  Parameters of the confluent Heun functions for an uncharged massless scalar field ($Q_{e}=Q_{m}=e=\mu_{0}=0$) in the conical Kerr metric, where $m_{(b)}=m/b$.}
\normalsize
\begin{tabular*}{170mm}{@{\extracolsep{\fill}}lll@{\extracolsep{\fill}}}
\toprule Parameter & Angular function $S(z)$                                             & Radial function $R(z)$                                                         \\\hline
			$\alpha$ & $4 a \omega$                                         & $2i\omega(r_{+}-r_{-})$                                         \\
			&& \\
			$\beta$  & $m_{(b)}$                                            & $2i\frac{ \omega(r_{+}^2+a^2) -a m_{(b)} }{ r_{+}-r_{-}}$       \\
			&& \\
			$\gamma$ & $m_{(b)}$                                            & $2i\frac{ \omega(r_{-}^2+a^2) -a m_{(b)} }{ r_{+}-r_{-}}$       \\
			&& \\
			$\delta$ & $0$                                                  & $2 \omega ^2(r_{+}+r_{-})(r_{-}-r_{+})$                         \\
			&& \\
			$\eta$   & $-\frac{4 a \omega m_{(b)}-m_{(b)}^2+2 \lambda }{2}$ & $-\frac{2 a^{2}[a \omega-m_{(b)}]^{2} +4a^2 \omega ^2 r_{+}r_{-}+(r_{+}-r_{-})^2 \lambda +4 a \omega m_{(b)} r_{+} r_{-}+2 r_{+}^3 \omega ^2 (r_{+}-2 r_{-})}{(r_{+}-r_{-})^2 }$ \\
\bottomrule
\end{tabular*}%
\end{center}

\begin{multicols}{2}

In what follows we will study the acoustic black hole analogue of the conical Kerr metric. Since Unruh \cite{PhysRevLett.46.1351,PhysRevD.51.2827} showed that the behavior of quantum systems in a classical gravitational field can be modeled by the motion of sound waves in a convergent fluid flow, the acoustic analogue of a black hole has been studied in the literature as a concrete laboratory model to test several aspects of the quantum field theory in curved spacetime \cite{Jacobson:2013,PhysRevD.91.104038,IntJModPhysD.26.1750035}.
%
%
\section{Analogue gravity: draining bathtub fluid flow with a disclination}
In this draining bathtub flow model, the velocity potential is given by \cite{ClassQuantumGrav.15.1767}
\begin{equation}
\Phi(r,\phi)=A \log r+B \phi\ ,
\label{eq:velocity_potential}
\end{equation}
where $A$ and $B$ are real constants. This leads to the following velocity of the fluid flow
\begin{equation}
\vec{v}=\frac{A}{r}\hat{r}+\frac{B}{r}\hat{\phi}\ .
\label{eq:velocity}
\end{equation}
Thus, the acoustic metric appropriate to a draining bathtub model (or rotating acoustic black hole), which is the analogue black hole metric (2+1)-dimensional with Lorentzian signature, has the line element given by
\begin{equation}
ds^{2}=-\frac{1}{r^{2}}\left(\Delta-\frac{B^{2}}{c^{2}}\right)dt^{2}+\frac{r^{2}}{\Delta}\ dr^{2}+r^{2}\ d\phi^{2}-2\frac{B}{c}\ d\phi\ dt\ ,
\label{eq:metrica_draining_bathtub_Kerr}
\end{equation}
with
\begin{equation}
\Delta=r^{2}-\frac{A^{2}}{c^{2}}=(r-r_{+})(r-r_{-})\ ,
\label{eq:Delta_draining_bathtub_Kerr}
\end{equation}
\begin{equation}
r_{\pm}=\pm r_{h}\ ,
\label{eq:sol_padrao_draining_bathtub_Kerr}
\end{equation}
\begin{equation}
r_{h}=\frac{|A|}{c}\ ,
\label{eq:horizon}
\end{equation}
where $c$ and $r_{h}$ are the speed of sound (constant throughout the fluid flow) and the acoustic event horizon (or Cauchy horizon of the rotating acoustic black hole), respectively. For $A < 0$ we are dealing with a future acoustic horizon, that is, an acoustic black hole.

We can introduce a disclination (topological defect) in the draining bathtub fluid flow (effective acoustic geometry). To do this, let us use a geometric approach and simply redefine the azimuthal angle $\phi$ in such a way that $\phi \rightarrow b\phi$. Then, we get the following line element in cylindrical coordinates \cite{Marques:2010}
\begin{equation}
ds^{2}=-\frac{1}{r^{2}}\left(\Delta-\frac{B^{2}}{c^{2}}\right)dt^{2}+\frac{r^{2}}{\Delta}\ dr^{2}+r^{2}b^{2}\ d\phi^{2}-2\frac{B}{c}b\ d\phi\ dt\ .
\label{eq:metrica_draining_bathtub_Kerr_disc}
\end{equation}
This metric has a deficit angle, which corresponds to removing $0 < b \leq 1$ or inserting $2\pi > b \geq 1$, a wedge of material of dihedral angle $\lambda=2\pi(b-1)$ by the Volterra process of disclination creation \cite{Kleman:1983}.
%
%
\section{Massless scalar field equation}
The equation of motion for the velocity potential describing a sound wave is identical to the general perturbation equation for the massless scalar field in curved spacetime \cite{LivingRevRelativity.8.12}, that is, to the covariant Klein-Gordon equation, which has the form
\begin{equation}
\frac{1}{\sqrt{-g}}\partial_{\sigma}\left(g^{\sigma\tau}\sqrt{-g}\partial_{\tau}\right)\Psi=0\ .
\label{eq:Klein-Gordon_cova_draining_bathtub_Kerr}
\end{equation}

Thus, the Klein-Gordon equation can be written in the spacetime (\ref{eq:metrica_draining_bathtub_Kerr_disc}) as
\begin{eqnarray}
&& \left[-\frac{r^{3}}{\Delta}\frac{\partial^{2}}{\partial t^{2}}+\frac{\partial}{\partial r}\left(\frac{\Delta}{r}\frac{\partial}{\partial r}\right)+\frac{1}{b^{2}r}\left(1-\frac{B^{2}}{c^{2}\Delta}\right)\frac{\partial^{2}}{\partial\phi^{2}}\right.\nonumber\\
&& -\left.\frac{2Br}{bc\Delta}\frac{\partial^{2}}{\partial\phi\ \partial t}\right]\Psi(\mathbf{r},t)=0\ .
\label{eq:mov_draining_bathtub_Kerr_disc}
\end{eqnarray}

We take the solution of Eq.~(\ref{eq:mov_draining_bathtub_Kerr_disc}) as follows
\begin{equation}
\Psi(\mathbf{r},t)=\sum_{m=-\infty}^{+\infty}R_{m\omega}(r)\mbox{e}^{im\phi}\mbox{e}^{-i \omega t}\ ,
\label{eq:separacao_variaveis_draining_bathtub_Kerr_disc}
\end{equation}
with $\omega$ being the frequency (or energy in natural units) of the particles where we assume that $0 < \omega \leq \infty$, and $m$ is a real constant that is not restricted to assume only a discrete set of values, because we are working with only two space dimensions. Substituting Eq.~(\ref{eq:separacao_variaveis_draining_bathtub_Kerr_disc}) into (\ref{eq:mov_draining_bathtub_Kerr_disc}), we find that the function $R_{m\omega,b}(r)$ satisfies the following equation

\end{multicols}

\begin{equation}
\biggr\{\frac{d}{dr}\left(\frac{\Delta}{r}\frac{d}{dr}\right)+\biggr[\frac{\omega^{2}r^{3}}{\Delta}-\frac{m_{(b)}^{2}}{r}\left(1-\frac{B^{2}}{c^{2}\Delta}\right)-\frac{2 B m_{(b)} \omega r}{c\Delta}\biggl]\biggl\}R_{m\omega,b}(r)=0\ ,
\label{eq:mov_radial_1_draining_bathtub_Kerr_disc}
\end{equation}

\begin{multicols}{2}

where $m_{(b)} \equiv m/b$.
%
%
\subsection{The radial equation}
In order to obtain the analytical solution of the radial Klein-Gordon equation, let us perform a change of variable such that
\begin{equation}
x=\frac{r^{2}}{2}\ .
\label{eq:x_draining_bathtub_Kerr_disc}
\end{equation}
Using this new coordinate, Eq.~(\ref{eq:Delta_draining_bathtub_Kerr}) can be rewritten as
\begin{equation}
\Delta=2x-\frac{A^{2}}{c^{2}}=2(x-x_{h})\ ,
\label{eq:Delta_x_draining_bathtub_Kerr_disc}
\end{equation}
where
\begin{equation}
x_{h}=\frac{1}{2}r_{h}^{2}
\label{eq:horizon_x_draining_bathtub_Kerr_disc}
\end{equation}
is the root of $\Delta$ and corresponds to the new acoustic event (and Cauchy) horizon of the background under consideration.

With this transformation, Eq.~(\ref{eq:mov_radial_1_draining_bathtub_Kerr_disc}) can be written as

\end{multicols}

\begin{eqnarray}
&& \frac{d^{2}R_{m\omega,b}(x)}{dx^{2}}+\left(\frac{1}{x-x_{h}}\right)\frac{dR_{m\omega,b}(x)}{dx}+\biggr[\frac{m_{(b)}^2 (B^2+2 c^2 x_{h})}{8 c^2 x_{h}^2}\frac{1}{x}+\frac{-B^2 m_{(b)}^2-2 c^2 m_{(b)}^2 x_{h}+4 c^2 x_{h}^2 \omega ^2}{8 c^2 x_{h}^2 }\frac{1}{x-x_{h}}\nonumber\\
&& +\frac{B^2 m_{(b)}^2-4 B c m_{(b)} x_{h} \omega +4 c^2 x_{h}^2 \omega ^2}{8 c^2 x_{h} }\frac{1}{(x-x_{h})^2}\biggl]R_{m\omega,b}(x)=0\ .
\label{eq:mov_radial_2_draining_bathtub_Kerr_disc}
\end{eqnarray}

\begin{multicols}{2}

Since this equation has singularities at $x=(x_{h},0,\infty)$, by the homographic substitution
\begin{equation}
z=\frac{x-x_{h}}{0-x_{h}}\ ,
\label{eq:z_draining_bathtub_Kerr_disc}
\end{equation}
we bring Eq.~(\ref{eq:mov_radial_2_draining_bathtub_Kerr_disc}) into Heun form as follows

\end{multicols}

\begin{eqnarray}
&& \frac{d^{2}R_{m\omega,b}(z)}{dz^{2}}+\frac{1}{z}\frac{dR_{m\omega,b}(z)}{dz}+\biggl\{\frac{B^2 m_{(b)}^2+2 c^2 m_{(b)}^2 x_{h}-4 c^2 x_{h}^2 \omega ^2}{8 c^2 x_{h}}\frac{1}{z}+\frac{-m_{(b)}^2 (B^2+2 c^2 x_{h})}{8 c^2 x_{h}}\frac{1}{z-1}\nonumber\\
&& -\biggl[i\left(\frac{2 c x_{h} \omega -m_{(b)} B}{2 c \sqrt{2x_{h}}}\right)\biggr]^{2}\frac{1}{z^2}\biggl\}R_{m\omega,b}(z)=0\ .
\label{eq:mov_radial_x_draining_bathtub_Kerr_disc_3}
\end{eqnarray}

\begin{multicols}{2}

Having thus moved the singularities to the points $z=0,1$, now we make the \textit{F-homotopic transformation} of the dependent variable $R_{m\omega,b}(z) \mapsto U_{m\omega,b}(z)$, namely
\begin{equation}
R_{m\omega,b}(z)=z^{A_{1}}U_{m\omega,b}(z)\ ,
\label{eq:s-homotopic_draining_bathtub_Kerr_disc_3}
\end{equation}
where the coefficient $A_{1}$ is given by
\begin{equation}
A_{1}=i\left(\frac{2 c x_{h} \omega -m_{(b)} B}{2 c \sqrt{2x_{h}}}\right)\ .
\label{eq:A1_draining_bathtub_Kerr_disc_3}
\end{equation}

Explicity, the result of applying (\ref{eq:s-homotopic_draining_bathtub_Kerr_disc_3}) to a radial equation in the form of (\ref{eq:mov_radial_x_draining_bathtub_Kerr_disc_3}) is
\begin{eqnarray}
&& \frac{d^{2}U_{m\omega,b}(z)}{dz^{2}}+\left(\frac{2A_{1}+1}{z}\right)\frac{dU_{m\omega,b}(z)}{dz}\nonumber\\
&& +\biggl[\frac{B^2 m_{(b)}^2+2 c^2 m_{(b)}^2 x_{h}-4 c^2 x_{h}^2 \omega ^2}{8 c^2 x_{h}}\frac{1}{z}\nonumber\\
&& -\frac{m_{(b)}^2 (B^2+2 c^2 x_{h})}{8 c^2 x_{h}}\frac{1}{z-1}\biggr]U_{m\omega,b}(z)=0\ .
\label{eq:mov_radial_x_draining_bathtub_Kerr_disc_4}
\end{eqnarray}

Thus, the linearly independent general exact solution of the radial Klein-Gordon equation for a massless scalar field in the rotating acoustic black hole with a disclination, over the entire range $0 \leq z < \infty$, can be written as
\begin{eqnarray}
R_{m\omega,b}(z) & = & z^{\frac{1}{2}\beta}\{C_{1_{m\omega,b}}\ \mbox{HeunC}(\alpha,\beta,\gamma,\delta,\eta;z)\nonumber\\
& + & C_{2_{m\omega,b}}\ z^{-\beta}\ \mbox{HeunC}(\alpha,-\beta,\gamma,\delta,\eta;z)\}\ ,\nonumber\\
\label{eq:solucao_geral_radial_draining_bathtub_Kerr_disc}
\end{eqnarray}
where $C_{1_{m\omega,b}}$ and $C_{2_{m\omega,b}}$ are constants to be determined, and the parameters $\alpha$, $\beta$, $\gamma$, $\delta$, and $\eta$ are given by
\begin{equation}
\alpha=0\ ,
\label{eq:alpha_radial_HeunC_draining_bathtub_Kerr_disc}
\end{equation}
\begin{equation}
\beta=i\left(\frac{2 c x_{h} \omega -m_{(b)} B}{c \sqrt{2x_{h}}}\right)\ ,
\label{eq:beta_radial_HeunC_draining_bathtub_Kerr_disc}
\end{equation}
\begin{equation}
\gamma=-1\ ,
\label{eq:gamma_radial_HeunC_draining_bathtub_Kerr_disc}
\end{equation}
\begin{equation}
\delta=-\frac{x_{h} \omega ^2}{2}\ ,
\label{eq:delta_radial_HeunC_draining_bathtub_Kerr_disc}
\end{equation}
\begin{equation}
\eta=\frac{1}{8} \left[m_{(b)}^2 \left(-\frac{B^2}{c^2 x_{h}}-2\right)+4 x_{h} \omega ^2+4\right]\ .
\label{eq:eta_radial_HeunC_draining_bathtub_Kerr_disc}
\end{equation}

Note the dependence of the radial solution on the parameter $b$, associated with the presence of a disclination.
%
%
\section{Hawking-Unruh radiation}
The radial solution given by Eq.~(\ref{eq:solucao_geral_radial_draining_bathtub_Kerr_disc}) has the following asymptotic behaviour at the exterior acoustic event horizon $r_{h}$:
\begin{equation}
R(r) \sim C_{1}\ (r-r_{h})^{\frac{1}{2}\beta}+C_{2}\ (r-r_{h})^{-\frac{1}{2}\beta}\ ,
\label{eq:exp_0_solucao_geral_radial_draining_bathtub_Kerr_disc}
\end{equation}
where all constants involved are included in $C_{1}$ and $C_{2}$.

From Eq.~(\ref{eq:beta_radial_HeunC_draining_bathtub_Kerr_disc}), the parameter $\beta$ can be written as
\begin{equation}
\beta=\frac{i}{\kappa_{h}}(\omega-\omega_{h,b})\ ,
\label{eq:expoente_rad_Hawking_draining_bathtub_Kerr_disc}
\end{equation}
with $\omega_{h,b}=m\Omega_{h,b}$, where the gravitational acceleration on the background acoustic horizon surface, $\kappa_{h}$, and the angular velocity of the exterior acoustic black hole in the presence of a disclination, $\Omega_{h,b}$, are given by
\begin{equation}
\kappa_{h} \equiv \frac{1}{2}\frac{1}{r_{h}^{2}}\left.\frac{d\Delta}{dr}\right|_{r=r_{h}}=\frac{1}{r_{h}}=\frac{c}{|A|}\ ,
\label{eq:acel_grav_ext_draining_bathtub_Kerr_disc}
\end{equation}
\begin{equation}
\Omega_{h,b}=\frac{Bc}{bA^{2}}\ .
\label{eq:vel_ang_draining_bathtub_Kerr_disc}
\end{equation}

Therefore, considering the time factor, on the acoustic black hole exterior horizon surface the ingoing and outgoing wave solutions are
\begin{equation}
\Psi_{in}=\mbox{e}^{-i \omega t}(r-r_{h})^{-\frac{i}{2\kappa_{h}}(\omega-\omega_{h,b})}\ ,
\label{eq:sol_in_1_draining_bathtub_Kerr_disc}
\end{equation}
\begin{equation}
\Psi_{out}(r>r_{h})=\mbox{e}^{-i \omega t}(r-r_{h})^{\frac{i}{2\kappa_{h}}(\omega-\omega_{h,b})}\ .
\label{eq:sol_out_2_draining_bathtub_Kerr_disc}
\end{equation}
These solutions depend on the parameter $b$, in such a way that the total energy of the radiated particles (the phonons) is decreased due to presence of a disclination.

Following the same procedure developed in our recent paper \cite{AnnPhys.350.14}, the relative scattering probability of the scalar wave at the acoustic event horizon surface, $\Gamma_{h}$, and the Hawking-Unruh radiation spectrum of scalar particles, $\left|N_{\omega}\right|^{2}$, respectively, are given by
\begin{equation}
\Gamma_{h}=\left|\frac{\Psi_{out}(r>r_{h})}{\Psi_{out}(r<r_{h})}\right|^{2}=\mbox{e}^{-\frac{2\pi}{\kappa_{h}}(\omega-\omega_{h,b})}\ ,
\label{eq:taxa_refl_draining_bathtub_Kerr_disc}
\end{equation}
\begin{equation}
\left|N_{\omega}\right|^{2}=\frac{1}{\mbox{e}^{\frac{2\pi}{\kappa_{h}}(\omega-\omega_{h,b})}-1}=\frac{1}{\mbox{e}^{\frac{\hbar}{k_{B}T_{h}}(\omega-\omega_{h,b})}-1}\ ,
\label{eq:espectro_rad_draining_bathtub_Kerr_disc_2}
\end{equation}
where $k_{B}T_{h}=\hbar\kappa_{h}/2\pi$ is the Hawking-Unruh radiation temperature, $k_{B}$ being the Boltzmann constant.

Therefore, we can see that the resulting Hawking-Unruh radiation spectrum of massless scalar particles has a thermal character, analogous to the black body spectrum. It is worth noticing that the total energy of radiated scalar particles is decreased due to the presence of a disclination. More precisely, the angular velocity of the exterior acoustic horizon, $\Omega_{h,b}$, is amplified in comparison with the scenario without a disclination \cite{GenRelativGravit.48.88}.
%
%
\section{Conclusions}
In this paper, we have obtained the exact and general solution for the radial part of the massless Klein-Gordon equation in the draining bathtub fluid flow with a disclination which corresponds to a rotating acoustic black hole in the presence of a topological defect. This model is an analogous gravity for the Kerr spacetime with a cosmic string.

This solution is analytic for all spacetime, which means, the region between the ergosphere and infinity. It is given in terms of the confluent Heun functions, and is valid over the range $0 \leq z < \infty$.

From the analytic solution corresponding to the radial part, we obtained the ingoing and outgoing scalar waves near the exterior acoustic horizon, and used these results to discuss the Hawking-Unruh radiation effect. As the angular velocity of the exterior acoustic black hole, $\Omega_{h,b}$, depends on the parameter $b$, this quantity codifies the influence of the disclination, and in fact, is amplified by the presence of this topological defect.

The wave function depends on the parameter $b$ that codifies the presence of a disclination, and as a consequence all other physical quantities are also influenced. Therefore, using this analogy, it is possible, in principle, to experimentally verify the Hawking-Unruh radiation emitted by acoustic black holes, and assuming that the physics which leads to Hawking radiation should be the same as its analogue, we can get some informations about the topological defect.

\vspace{15mm}

%
%
\acknowledgments{The author would like to thank Prof. V. B. Bezerra for the fruitful discussions.}

\end{multicols}

\vspace{-1mm}
\centerline{\rule{80mm}{0.1pt}}
\vspace{2mm}

\begin{multicols}{2}

%
%

\end{multicols}

\clearpage
\end{CJK*}

\begin{thebibliography}{99}

\vspace{3mm}

\bibitem{PhysLettB.686.279} R. Banerjee, B. R. Majhi and E. C. Vagenas, Phys. Lett. B \textbf{686}, 279 (2010).
\bibitem{PhysRevD.84.127502} P. Fiziev and D. Staicova, Phys. Rev. D \textbf{84}, 127502 (2011).
\bibitem{PhysRevD.87.084047} T. Jacobson and A. Satz, Phys. Rev. D \textbf{87}, 084047 (2013).
\bibitem{ClassQuantumGrav.31.045003} V. B. Bezerra, H. S. Vieira and A. A. Costa, Class. Quantum Grav. \textbf{31}, 045003 (2014).
\bibitem{CommunMathPhys.43.199} S. W. Hawking, Commun. Math. Phys. \textbf{43}, 199 (1975).
%
\bibitem{PhysRevD.82.044013} R. Banerjee, C. Kiefer and B. R. Majhi, Phys. Rev. D \textbf{82}, 044013 (2010)
\bibitem{PhysLettB.692.61} K. Umetsu, Phys. Lett. B \textbf{692}, 61 (2010).
\bibitem{PhysLettB.697.398} A. Yale, Phys. Lett. B \textbf{697}, 398 (2011).
\bibitem{EurophysLett.109.60006} H. S. Vieira, V. B. Bezerra and A. A. Costa, Europhys. Lett. \textbf{109}, 60006 (2015).
%
\bibitem{PhysRevD.83.124016} A. Fabbri and C. Mayoral, Phys. Rev. D \textbf{83}, 124016 (2011).
\bibitem{PhysRevD.83.124047} C. Mayoral, A. Fabbri and M. Rinaldi, Phys. Rev. D \textbf{83}, 124047 (2011).
\bibitem{NewJPhys.10.103001} I. Carusotto, S. Fagnocchi, A. Recati, R. Balbinot and A. Fabbri, New J. Phys. \textbf{10}, 103001 (2011).
\bibitem{PhysRevD.92.024043} J. Steinhauer, Phys. Rev. D \textbf{92}, 024043 (2015).
\bibitem{ClassQuantumGrav.18.1137} C. Barcel\'{o}, S. Liberati and M. Visser, Class. Quantum Grav. \textbf{18}, 1137 (2001).
\bibitem{GenRelGrav.34.1719} M. Visser, C. Barcel\'{o} and S. Liberati, Gen. Rel. Grav. \textbf{34}, 1719 (2002).
\bibitem{JHighEnergyPhys.04.030} S. R. Das, A. Ghosh, J. H. Oh and A. D. Shapere, J. High Energy Phys. \textbf{04}, 030 (2011).
\bibitem{JPhysBAtMolOptPhys.45.163001} S. J. Robertson, J. Phys. B: At. Mol. Opt. Phys. \textbf{45}, 163001 (2012).
\bibitem{PhysRevA.78.021601} S. W\"{u}ster, Phys. Rev. A \textbf{78}, 021601(R) (2008).
\bibitem{PhysRevD.90.104015} A. Belenchia, S. Liberati and A. Mohd, Phys. Rev. D \textbf{90}, 104015 (2014).
\bibitem{PhysRevD.81.124013} E. S. Oliveira, S. R. Dolan and L. C. B. Crispino, Phys. Rev. D \textbf{81}, 124013 (2010).
\bibitem{PhysRevD.79.064014} S. R. Dolan, E. S. Oliveira and L. C. B. Crispino, Phys. Rev. D \textbf{79}, 064014 (2009).
\bibitem{PhysRevD.70.124006} E. Berti, V. Cardoso and J. P. S. Lemos, Phys. Rev. D \textbf{70}, 124006 (2004).
\bibitem{PhysRevD.44.1731} T. Jacobson, Phys. Rev. D \textbf{44}, 1731 (1991).
\bibitem{PhysRevD.48.728} T. Jacobson, Phys. Rev. D \textbf{48}, 728 (1993).
%
\bibitem{Vilenkin:1994} A. Vilenkin and E. P. S. Shellard, \textit{Cosmic strings and other topological defects}, (Cambridge University Press, Cambridge, 1994).
\bibitem{ClassQuantumGrav.6.1313} D. V. Gal'tsov and E. Mas\'{a}r, Class. Quantum Grav. \textbf{6}, 1313 (1989).
\bibitem{Marques:2010} F. A. Gomes and G. A. Marques, in \textit{Astronomy and Relativistic Astrophysics}, edited by C. A. Z. Vasconcellos et al. (World Scientific, Singapore, 2010), pp.~153--159.
\bibitem{AnnPhys.362.576} H. S. Vieira, V. B. Bezerra and G. V. Silva, Ann. Phys. (NY) \textbf{362}, 576 (2015).
\bibitem{Ronveaux:1995} A. Ronveaux, \textit{Heun's differential equations}, (Oxford University Press, New York, 1995).
\bibitem{Slavyanov:2000} S. Y. Slavyanov and W. Lay, \textit{Special functions}, (Oxford University Press, New York, 2000).
\bibitem{PhysRevLett.46.1351} W. G. Unruh, Phys. Rev. Lett. \textbf{46}, 1351 (1981).
\bibitem{PhysRevD.51.2827} W. G. Unruh, Phys. Rev. D \textbf{51}, 2827 (1995).
\bibitem{Jacobson:2013} T. Jacobson, in \textit{Analogue gravity phenomenology}, lecture notes in physics, edited by D. Faccio et al. (Springer International Publishing, Switzerland, 2013), Vol. 870, pp.~1--29.
\bibitem{PhysRevD.91.104038} C. L. Benone, L. C. B. Crispino, C. Herdeiro and E. Radu, Phys. Rev. D \textbf{91}, 104038 (2015).
\bibitem{IntJModPhysD.26.1750035} H. S. Vieira, Int. J. Mod. Phys. D \textbf{26}, 1750035 (2017).
\bibitem{ClassQuantumGrav.15.1767} M. Visser, Class. Quantum Grav. \textbf{15}, 1767 (1998).
\bibitem{Kleman:1983} M. Kl\'{e}man, \textit{Points, lines and walls}, (Wiley, New York, 1983).
\bibitem{LivingRevRelativity.8.12} C. Barcel\'{o}, S. Liberati and M. Visser, Living Rev. Relativity \textbf{8}, 12 (2005).
\bibitem{AnnPhys.350.14} H. S. Vieira, V. B. Bezerra and C. R. Muniz, Ann. Phys. (NY) \textbf{350}, 14 (2014).
\bibitem{GenRelativGravit.48.88} H. S. Vieira and V. B. Bezerra, Gen. Relativ. Gravit. \textbf{48}, 88 (2016).
\end{thebibliography}
\end{document}